\documentclass[a4paper,11pt]{article}
\usepackage{pos}
\newcommand{\bra}[1]{\ensuremath{\langle #1 |}}
\newcommand{\ket}[1]{\ensuremath{| #1 \rangle}}
\newcommand{\braket}[2]{\ensuremath{\langle #1 | #2 \rangle}}

\title{Bootstrap methods for digitized scalar field theory}
\author*[a]{Zane Ozzello}
\author[a]{Yannick Meurice}

\affiliation[a]{ Department of Physics and Astronomy, The University of Iowa,\\
  Iowa City, IA 52242, USA}

\emailAdd{zane-ozzello@uiowa.edu}
\emailAdd{yannick-meurice@uiowa.edu}

\abstract{General positivity constraints linking various powers of observables in energy eigenstates can be used to sharply locate acceptable regions for the energy eigenvalues, provided that efficient recursive methods are available to calculate the matrix elements. These recursive methods are derived by looking at the commutation relations of the observables with the Hamiltonian. We discuss how this self-consistent (bootstrap) approach can be applied to the study of digitized scalar field theory in the harmonic basis. Using known results, we develop the method by testing on quantum systems, including the harmonic and anharmonic oscillators. We report recent numerical results for up to four coupled anharmonic oscillators. From here, we consider the possibility of using the groundwork of this method as a means of studying phase transitions in 1+1 dimensions.}

\FullConference{The 40th International Symposium on Lattice Field Theory (Lattice 2023)\\
July 31st - August 4th, 2023\\
Fermi National Accelerator Laboratory\\}

\begin{document}
\maketitle

\section{Introduction}
Universal quantum computers allow for real time evolution in quantum field theory.  One of the simplest examples is $\lambda \phi^4$ [1,2,3].  In [4], a digitization inspired  by Gaussian quadrature for the path integral where $\phi$ is restricted to the zeros of the $n_{max}$ order Hermite polynomial is used. With the following modification and resulting commutation relations, the Dyson Interaction picture is preserved:
        
\begin{equation}           
                 [a, a^\dagger] = {\bf{1}} - n_{max} \ket{n_{max}-1} \bra{n_{max}-1}
                \end{equation}
                \begin{equation}
                 [a^\dagger a,a]=-a
                 \end{equation}
                 \begin{equation}
                 [a^\dagger a, a^\dagger] = a^\dagger
\end{equation}           
It is shown in [4] that since digitization provides a field cutoff, the perturbation series in $\lambda$ will converge with a radius determined from complex singularities, for sufficiently small $|\lambda|$ [5].  See below for case when $n_{max}=8$ (from [4]).
        
        \begin{center}
            \includegraphics[width=4in,angle=0]{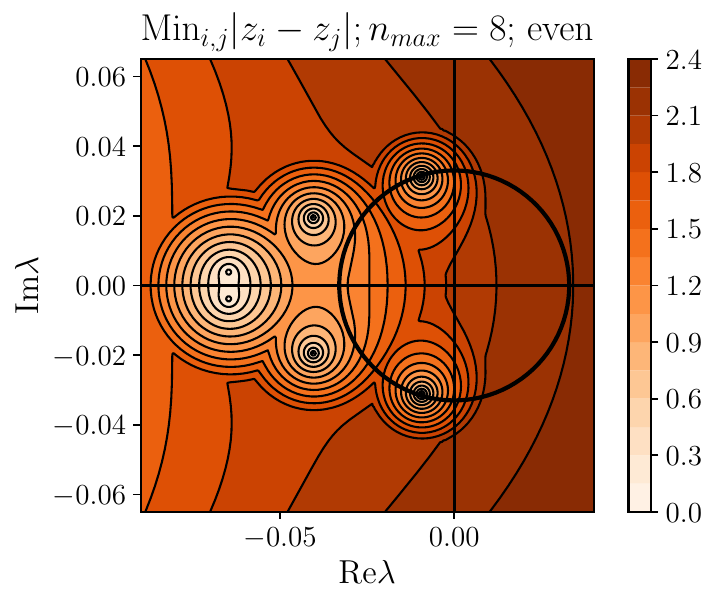}

 \scriptsize{\textbf{Figure 1: } From [4], an example of radius of convergence for $n_{max}=8$}
 \end{center}
In 1+1 dimensions these singularities pinch the real axis for $\lambda = \lambda_c$, which indicates second order phase transition.  

For the undigitized anharmonic oscillator, the moments $\bra{E} x^{2l} \ket{E} $ can be recursively calculated [6,7].

Within this work, we begin with a discussion of the methods that will yield this recursive process, the quantum mechanical bootstrap.  We then investigate a specific case for the infinite undigitized anharmonic oscillator, to allow us to eventually compare to results from applying the bootstrap to the digitized anharmonic oscillator.  Finally, we investigate what can be learned from combining the bootstrap with the digitization procedure.  

\clearpage

\section{Background}

\subsection{Bootstrap Foundations}
Utilizing the following simple identities and constraints with our Hamiltonian of interest, we can begin to develop a bootstrap.  
            \begin{equation} 
    		\langle [H,O] \rangle = 0
    	\end{equation}
    	\begin{equation}
    		\langle HO \rangle = E \langle O \rangle
    	\end{equation}
     	\begin{equation} 
    		\langle O^{\dagger}O \rangle \ge 0
    	\end{equation} 
All of these should be familiar from an introductory quantum mechanics course, but by combining the above equations with different momentum and positional moments, we begin to build the requisite background.  As is shown in [7], by utilizing $O= x^m$, $x^m p$, and $x^{m-1}$ we get, respectively: 
\begin{equation}
  s \langle x^{m-1}p \rangle = \frac{i}{2} m (m-1) \langle x^{m-2} \rangle
\end{equation}

\begin{equation}
    0=m \langle x^{m-1} p^2 \rangle+ \frac{1}{4} m (m-1)(m-2)\langle x^{m-3}\rangle - \langle x^m V'(x)\rangle
\end{equation}
\begin{equation}
     E \langle x^{m-1} \rangle = \frac{1}{2} \langle x^{m-1}p^2 \rangle + \langle x^{m-1} V(x) \rangle
\end{equation}
Upon combining the above equations, we recover an equation which can be used in the bootstrapping algorithm itself:
\begin{equation}
               0= 2mE\langle x^{m-1} \rangle   
             + \frac{1}{4}m(m-1)(m-2) \langle x^{m-3} \rangle 
             - \langle x^{m}V'(x) \rangle-2m \langle x^{m-1}V(x) \rangle 
\end{equation}
This is our moment recursion.  Utilizing this we can find higher moments of x in terms of lower moments, our potential, and our energy.  Notice, that upon setting m=1 this returns the virial theorem.  From the positivity of the norm, a Hankel matrix can be constructed as such:
 \begin{equation}
                0\le \langle O^{\dagger}O \rangle = \sum_{ij} c_i^* \langle x^{i+j} \rangle c_j \equiv \sum_{ij}c_i^*M_{ij}c_j       
            \end{equation}
We now have the necessary pieces for implementing the bootstrap algorithm.  

\subsection{Bootstrap Algorithm}
Starting with a Hamiltonian with a given potential, find a recursive statement for this potential with equation (10).  This will allows us to populate the Hankel matrix with moments according to (11).  Taking a search space S and a set of trial points A within S, we then utilize the algorithm by the following steps:
     \begin{enumerate}
        \item For each point in A, generate 2K-2 points of the moment sequence for each.  
        \item For these terms, construct a $K \times K$ Hankel matrix where $M_{ij} = \langle x^{i+j} \rangle$ for each point in A.
        \item Check positive definiteness of this matrix.  If positive definite accept this point, if not throw it out.  
        \item Obtain a set of values at depth K: $B_K$. Note: $B_K \subseteq A \subset S $ and, working through different values of K: $B_{K+1} \subseteq B_K$
    \end{enumerate}
This method will allow us to trial over different values at once, for example in the infinite anharmonic oscillator, we simultaneously trial over possible energies and possible $\langle x^2 \rangle$ moments, as these are needed to be provided for the moment recursion, as opposed to being found by it.  The ability to deal with multiple trial spaces simultaneously allows for the handling of more complicated Hamiltonians with the method.  

\section{Infinite Anharmonic Oscillator}
To see how the quantum mechanical bootstrap of [6,7] behaves, we tested the cases of the infinite harmonic oscillator and infinite anharmonic oscillator.  Results for the infinite harmonic oscillator were not illuminating as to the behavior of the algorithm, but by replicating the work in [6] for the infinite anharmonic oscillator a useful visualization of the algorithm at work arises.  

We use a Hamiltonian equation of the form $H = p^2 +gx^2 + hx^4 $ with $g=h=m=\omega=\hbar=1$

\begin{center}
 \includegraphics[width=1.9in, angle=0]{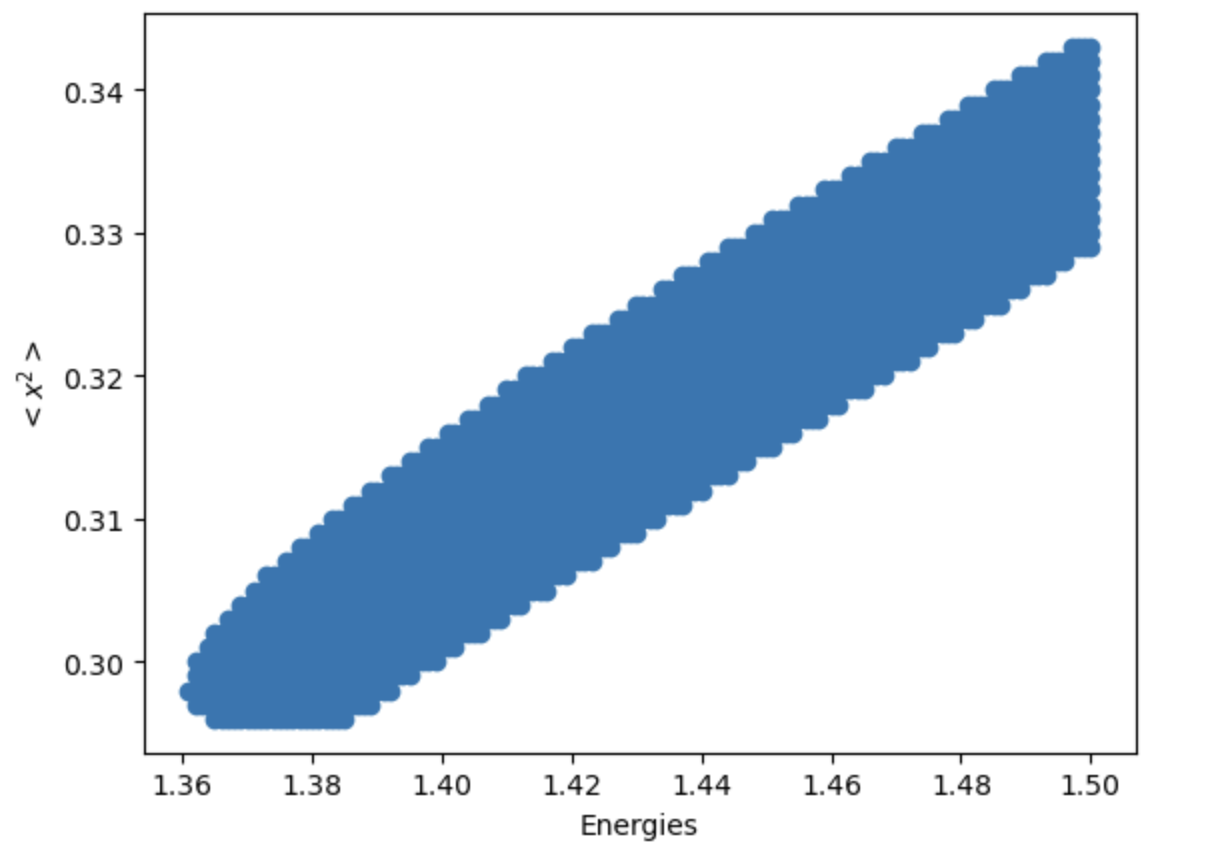}
 \includegraphics[width=1.9in,angle=0]{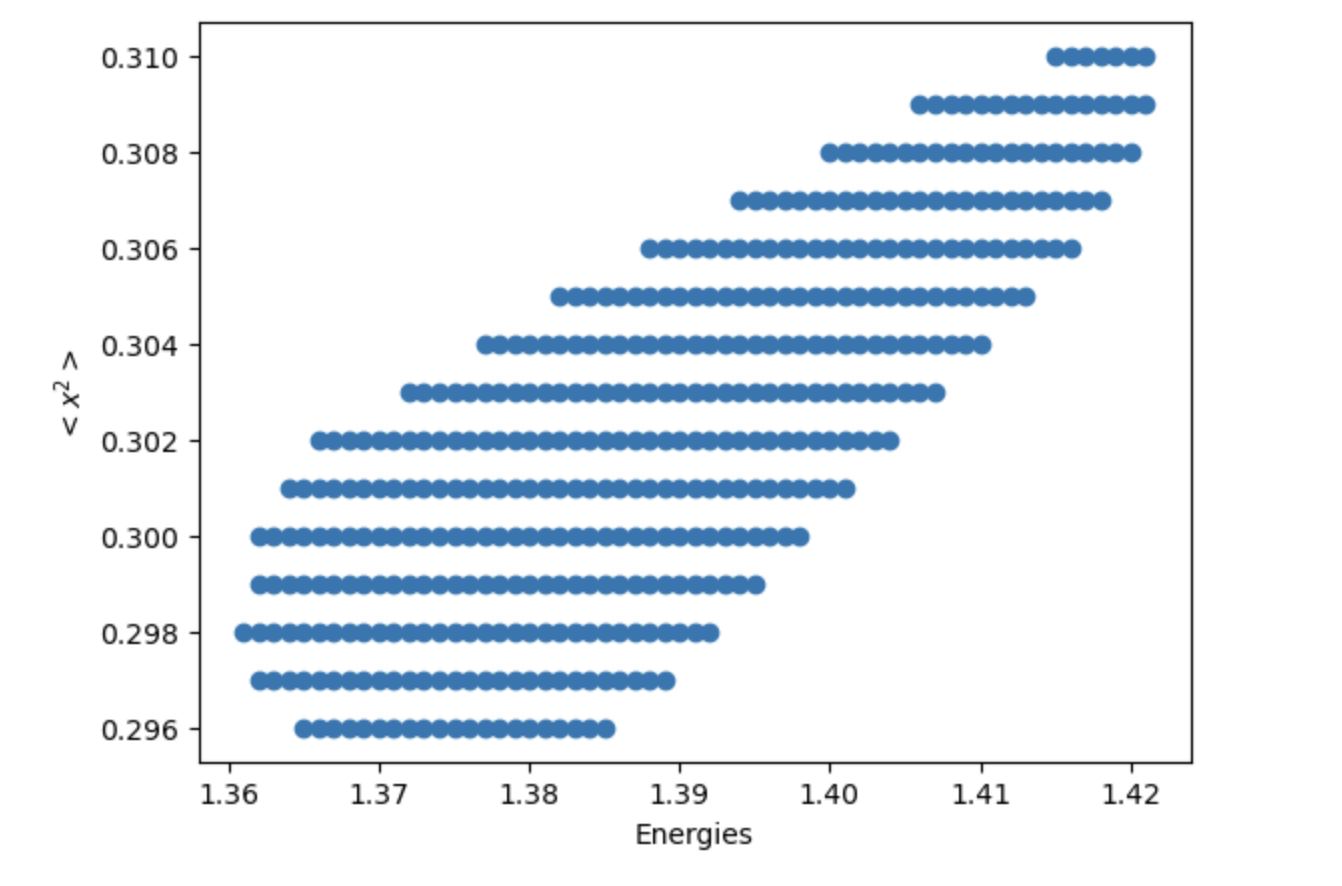}
%\vskip-0.5in
 \includegraphics[width=1.9in,angle=0]{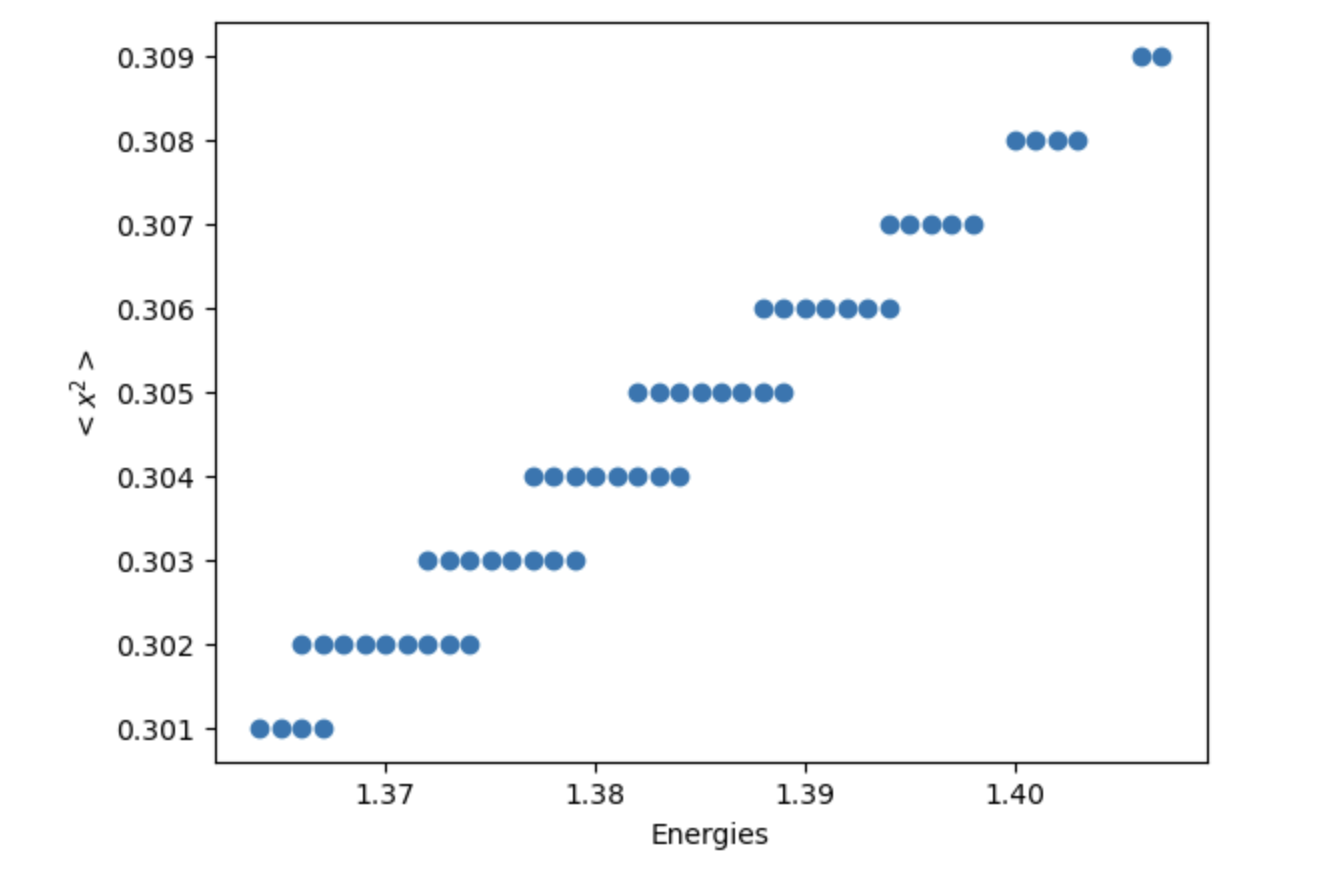}
 \includegraphics[width=1.9in,angle=0]{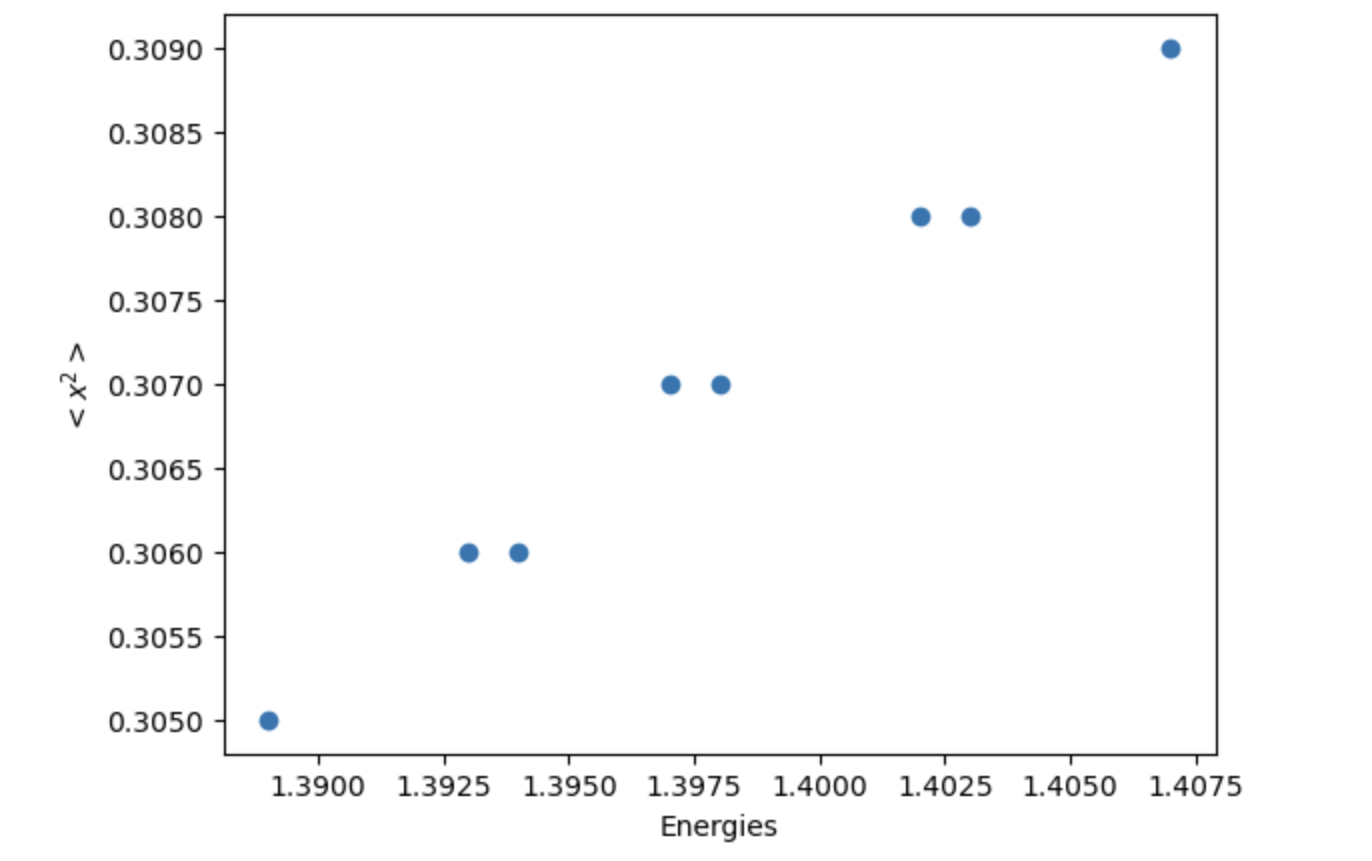}
%\vskip-0.5in $\ $
\includegraphics[width=1.9in,angle=0]{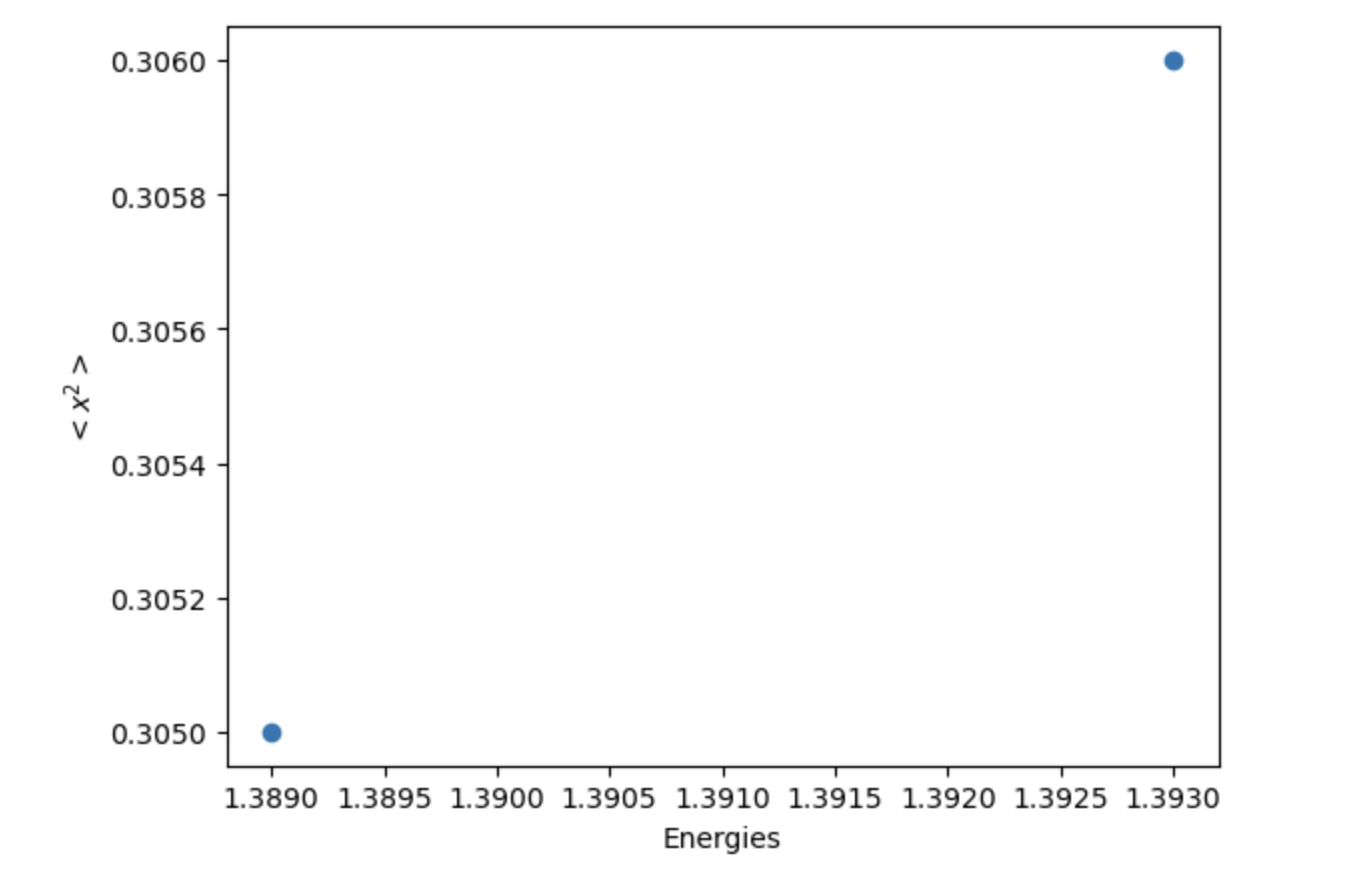}
 \includegraphics[width=1.9in,angle=0]{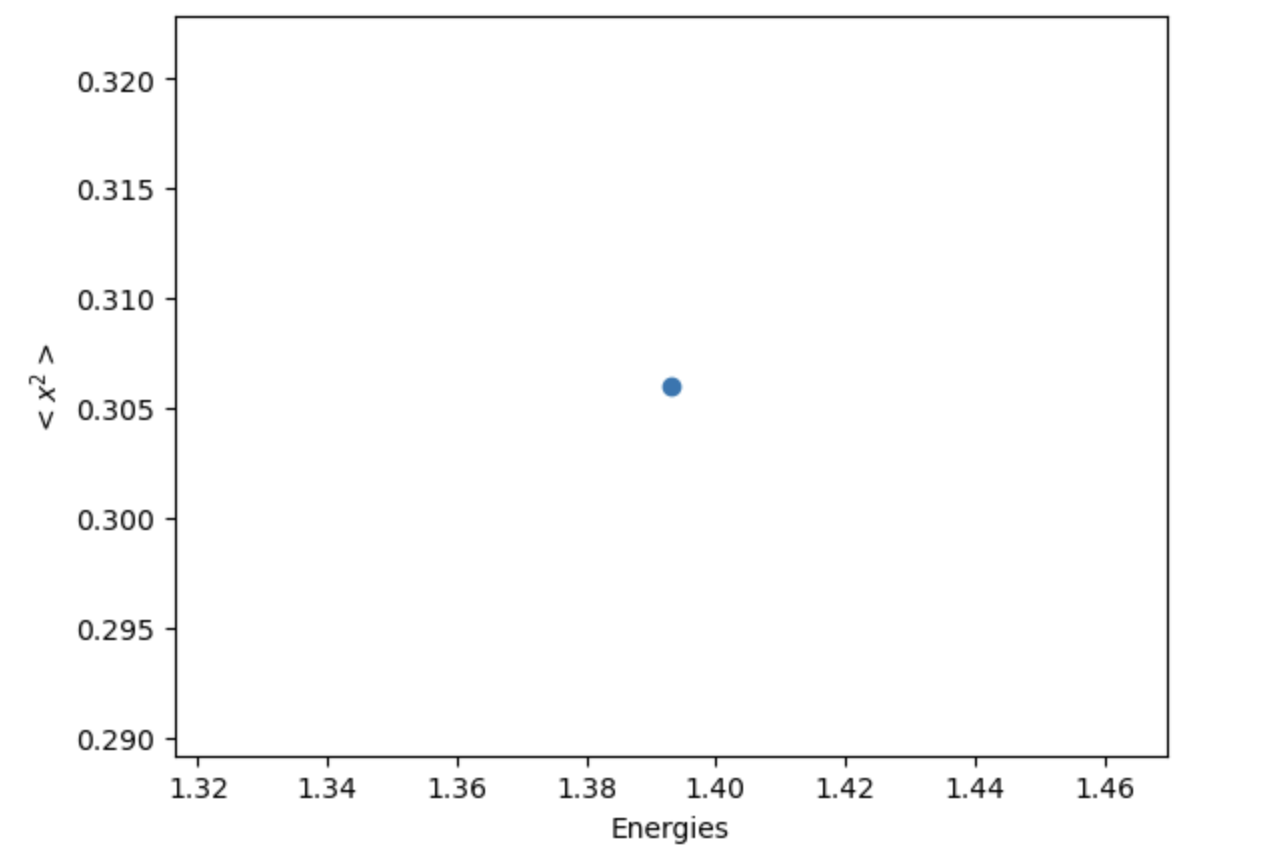}
 \end{center}

 \begin{center}
 \scriptsize{ \textbf{Figure 2: } From left to right, depths at K= 7,8,9,10,11,12 for the ground state energy}
 \end{center}
This highlights how the bootstrap behaves visually.  The blue regions are all possible specific $\langle x^2 \rangle$ and $E$ combinations for a given depth.  Recall, these depths for a $K \times K$ matrix tell us the highest moment used for depth K is $\langle x^{2k} \rangle$.  So, upon increasing the depth K, we reduce the possible $\langle x^2 \rangle$ and $E$ combinations (seen visually as the "islands" decreasing in size), eventually reducing to the energy values for the Hamiltonian.  

\section{Applications for Digitization}
We now seek to determine if the bootstrap method can be applied to the digitized case, and if so, what, if any modifications need to be made to implement.  From here, we investigate what information can be learned from such applications.  
\subsection{Digitized Oscillators}
Now, by taking $H_{n_{max}}(\hat{x})=0$ we can re-express the $\langle x^{n_{max}} \rangle$ moment in terms of lower powers of x, via the Hermite polynomial.  Recall, the Hermite polynomial is the eigenfunction for the harmonic oscillator, so by setting it to 0 a highest level is imposed at which this and all successive Hermite polynomials go to zero.  

From here, the $n_{max}=4$ case will be investigated.  From $H_4 (\hat{x}) = 0$, we get a rescursion relation:
            \begin{equation}
                \hat{x}^4 = 3 \hat{x}^2 - \frac{3}{4}
            \end{equation}
Which we can now use as we did a derived statement from (10) before.  Now, moments larger than $n_{max}=4$ are superfluous. This limits us to a $\frac{n_{max}}{2} \times \frac{n_{max}}{2}$ Hankel matrix, as we only need to go to moments of the $n_{max}=4$ power, and the odd moments are all zero, allowing us to reduce to this $2 \times 2$ matrix.  It is important to note that as a result of this truncation, these higher moments are not only superfluous, but entirely useless as the truncation removes their relevance.  This digitization (1,2,3) introduces new terms as a result of the modifications to the relations, yielding new terms are introduced from the modified relations, for instance in the $n_{max}=4$ case terms such as $\braket{E}{n_{max}-1}$.

Knowing $\bra{E} \hat{x}^0 \ket{E} , \bra{E} \hat{x}^2 \ket{E}, ..., \bra{E} \hat{x}^{n_{max}-2} \ket{E} $ is equivalent to the knowledge of $|\braket{x_j}{E} |^2 \ge 0$ for the $\frac{n_{max}}{2}$ positive zeros of the Hermite polynomial of degree $n_{max}$.  We predict that the positivity constraint imposed by the eigenvalues of the $\frac{n_{max}}{2} \times \frac{n_{max}}{2}$ Hankel matrix as a result of the positive definiteness are equivalent to the constraints on energy also imposed by the necessary positivity of $|\braket{x_j}{E} |^2 \ge 0$ as a probability.  Specifically, we will attempt to show this for a digitized harmonic and anharmonic oscillator with $n_{max} = 4$.   

\subsection{Digitized Results}
When producing the bootstrap for each case, instead of trialing over sample $\langle x^2 \rangle$ values randomly, utilizing a position operator built from raising and lowering operators, we can provide $\langle x^2 \rangle$ in terms of energy to be used directly, that will differ with the Hamiltonian.  

Resultingly, the digitized harmonic oscillator yields:  
            \begin{center}
            \includegraphics[width=4in,angle=0]{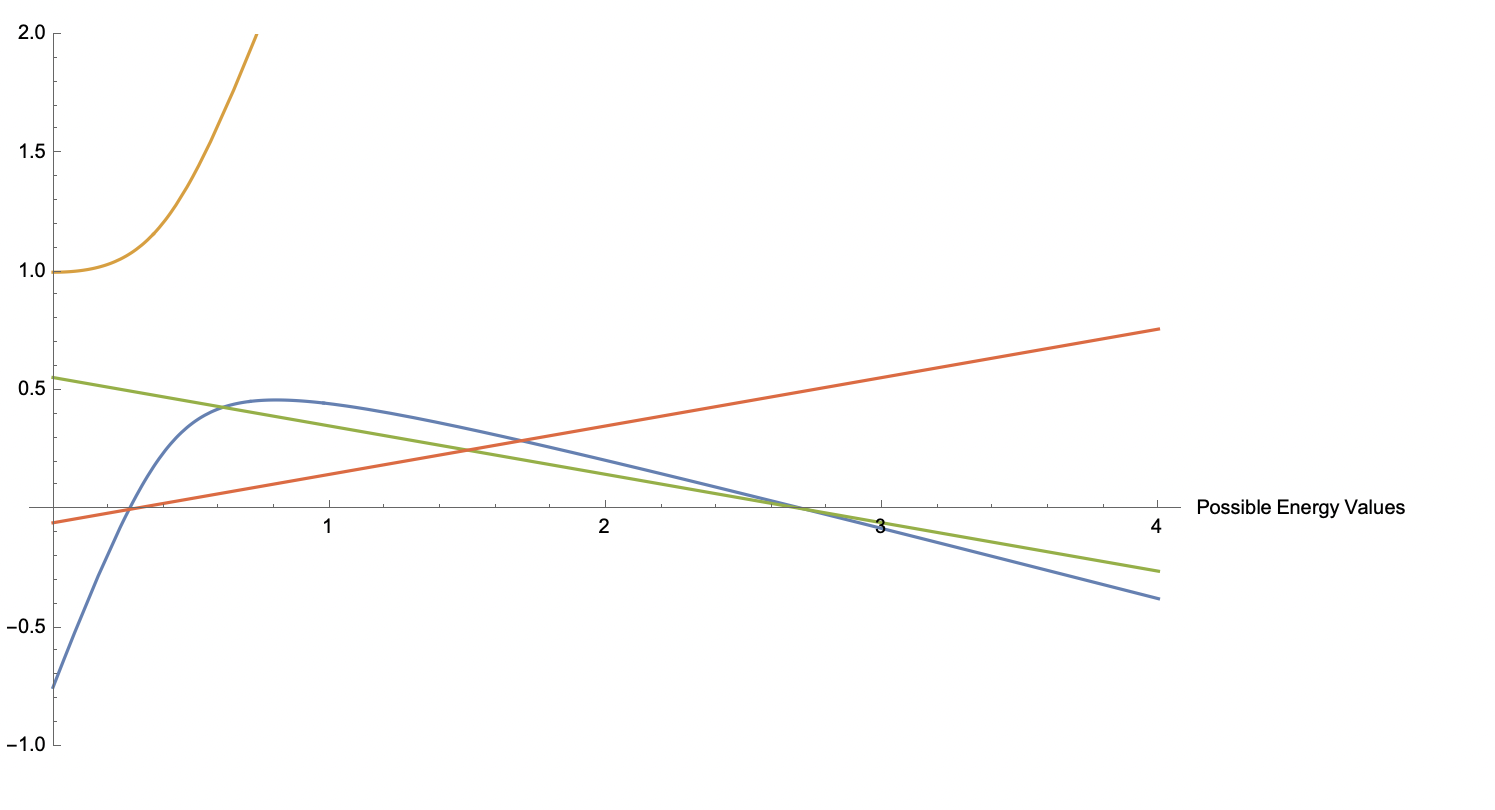}
            \end{center}
    \begin{center}
    \scriptsize{ \textbf{Figure 3: } Graph of energy constraints imposed by eigenvalues (orange and blue) and those imposed by Hermite polynomials (red and green)}
    \end{center}

It can be seen on the x-axis we get matching intercepts for results from our probability and from our eigenvalue constraints at the same spots, yielding an area of possible values together.  Energy falls within the range $\frac{3-\sqrt{6}}{2}$ to $\frac{3+\sqrt{6}}{2}$. Here we have:
            
            \begin{equation}
                \langle x^2 \rangle = E- 2|\braket{E}{3}|^2
            \end{equation}
With the inner product set to 0 for convenience, but if included would just provide a shift of the graph.  

When looking at the digitized anharmonic oscillator we get:
          
            \begin{equation}
                \langle x^2 \rangle = \frac{1}{6\lambda +1}(E- 2|\braket{E}{3}|^2 
                      + 3\lambda \sqrt{6} \braket{E}{1}\braket{3}{E} 
                      +9\lambda |\braket{E}{2}|^2+9\lambda |\braket{E}{3}|^2)
            \end{equation}
One can see how with $\lambda=0$ this returns the harmonic case.  Which gives for the digitized anharmonic oscillator: 
\begin{center}
\includegraphics[width=2.5in,angle=0]{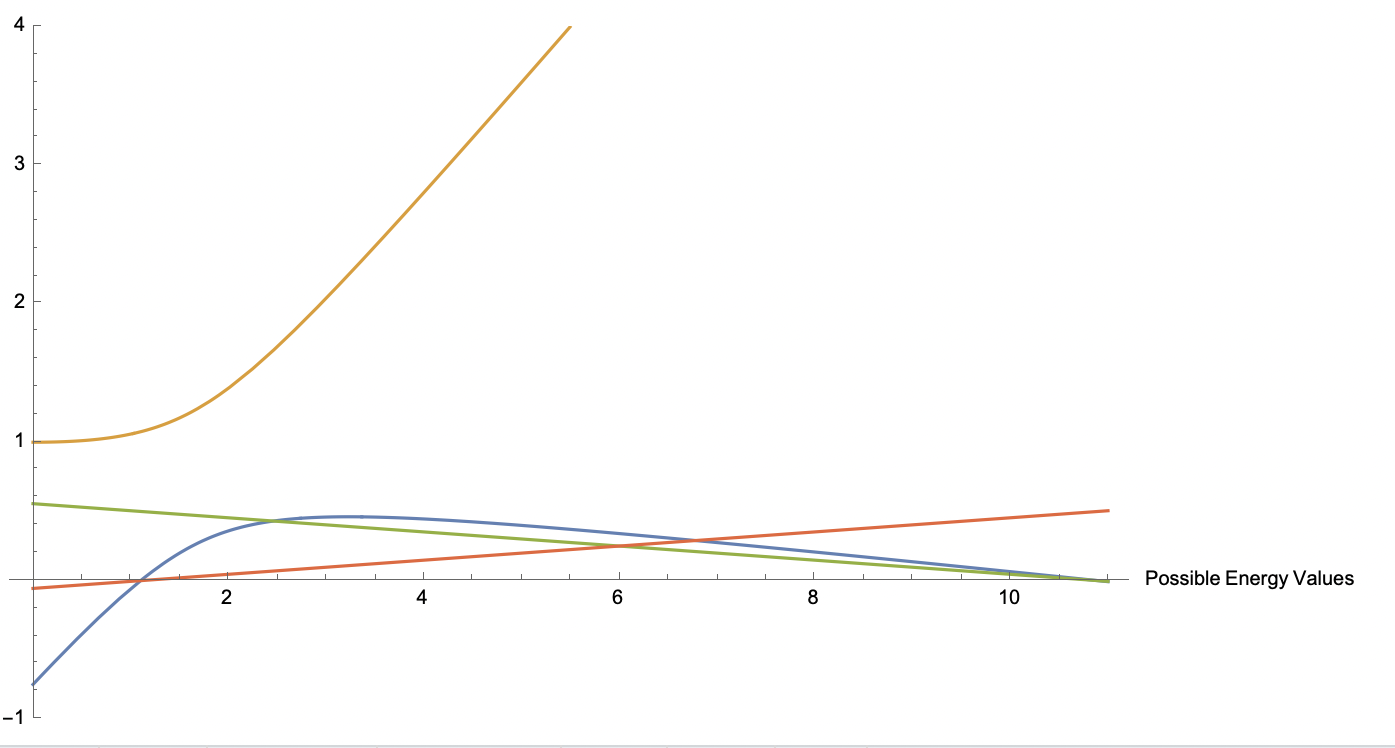}
\includegraphics[width=2.5in,angle=0]{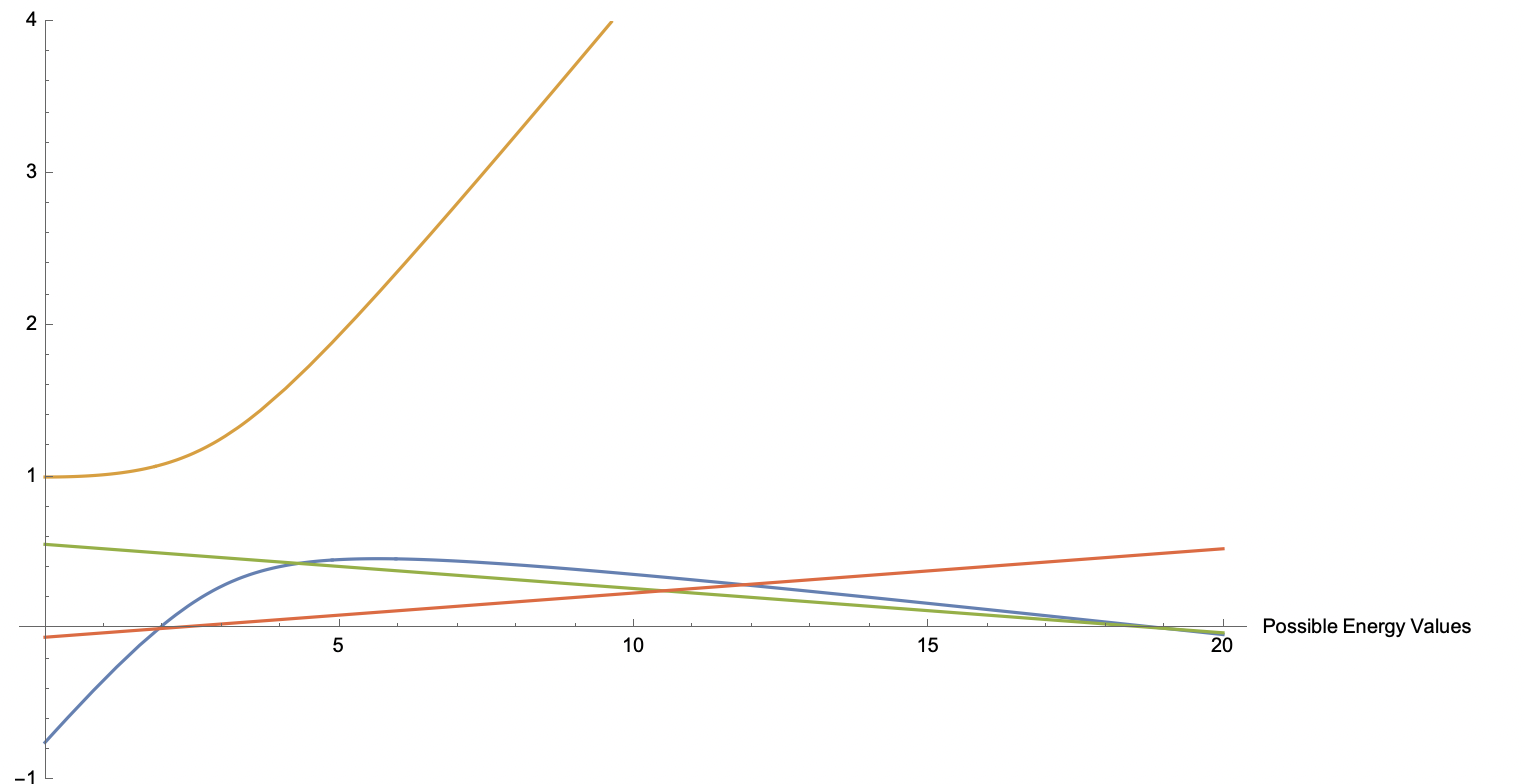}
\end{center}
    \begin{center}
    \scriptsize{ \textbf{Figure 4: } Left figure is $\lambda =\frac{1}{2}$ and right is $\lambda = 1$ Graph of energy constraints imposed by eigenvalues (orange and blue) and those imposed by Hermite polynomials (red and green)}
    \end{center}

Noticing, these graphs are not exceptionally different from the harmonic case.  The anharmonic case just introduces a scaling difference with a coordinate shift of the same graph.  For $\lambda = \frac{1}{2}$ and $\lambda = 1$ they would respectively have energy fall in ranges of $6 - 2\sqrt{6}$ to $6 + 2\sqrt{6}$ for the former and $\frac{21 - 7\sqrt{6}}{2}$ to $\frac{21 + 7\sqrt{6}}{2}$ for the latter. 

\section{Future Work}
We have applied the quantum mechanical bootstrap to different oscillator models.  We have shown that with minimal adjustments, the bootstrap can be applied to the digitized case, showing for the harmonic and anharmonic case for $n_{max}=4$.  Going forward, we want to see how easily we can apply this method to larger $n_{max}$, with beginning to look at $n_{max}=8$.  We also want to investigate how the bootstrap works with coupled oscillators. 

\section*{Acknowledgements}
This work is supported in part by the Department of Energy under
Award Numbers DE-SC0019139 and DE-SC0010113.

\end{document}